\documentclass{article}
\usepackage{spconf,amsmath, amssymb}

\usepackage[pdftex]{graphicx}
\usepackage{multirow}
\usepackage{here}
\usepackage{multicol}
\DeclareMathOperator{\E}{\mathbb{E}}

\newcommand{\bhline}[1]{\noalign{\hrule height #1}}  

\title{Disentangling physical parameters for anomalous sound detection under domain shifts}
\name{Kota Dohi, Takashi Endo, Yohei Kawaguchi}

\address{Research and Development Group, Hitachi, Ltd.\\
1-280, Higashi-koigakubo, Kokubunji-shi, Tokyo 185-8601, Japan\\
\texttt{kota.dohi.gr@hitachi.com}}

\begin{document}
\ninept
\maketitle
\begin{abstract}\vspace*{-2.8mm}
  To develop a sound-monitoring system for machines, a method for detecting 
  anomalous sound under domain shifts is proposed.
  A domain shift occurs when a machine's physical parameters change. Because a domain shift 
  changes the distribution of normal sound data, conventional 
  unsupervised anomaly detection methods can output false positives.
  To solve this problem, the proposed method constrains some latent variables of 
  a normalizing flows (NF) model to represent physical parameters, which 
  enables disentanglement of the factors of domain shifts and learning of  
  a latent space that is invariant with respect to these domain shifts.
  Anomaly scores calculated from this domain-shift-invariant latent space
  are unaffected by such shifts, which reduces false positives
  and improves the detection performance.
  Experiments were conducted with sound data from a slide rail under different 
  operation velocities.
  The results show that the proposed method disentangled the velocity to 
  obtain a latent space that was invariant with respect to domain shifts, which improved the AUC 
  by 13.2\% for Glow with a single block and 2.6\% for Glow with multiple blocks.                                      
\end{abstract}
\begin{keywords}
Machine health monitoring, Anomalous sound detection, Anomaly detection, 
Semi-supervised disentanglement learning, 
Normalizing flows
\end{keywords}

\section{Introduction}\vspace*{-2.8mm}
\label{sec:intro}
As the number of skilled maintenance workers decreases worldwide, the demand for automatic 
sound-monitoring system has been increasing.
Such systems aim to detect anomalous states of a machine from its sound.

Because anomalous sound data can rarely be obtained in practice, unsupervised anomaly detection 
methods are often adopted for these systems \cite{Suefusa2020,Koizumi2020}.
Neural generative models such as a variational autoencoder (VAE) \cite{Kingma2014a} and a normalizing flows (NF) \cite{Tabak2013,Dinh2015}
are the most commonly used methods for unsupervised anomaly detection 
because of their high detection performance.
These methods try to detect data with different distributions from normal data, without 
using anomalous data.
However, not only a machine's anomalous states but also changes in its physical parameters 
(domain shifts) or aging can affect the distributions of the normal data, which induces
false positives when using unsupervised methods.
Aging causes changes in data distributions over a long period of time,  
and these changes can be handled by continual learning or incremental learning \cite{Wiewel2019}.
On the other hand, domain shifts, which are the focus in this paper, 
can induce sudden, huge differences in data 
distributions, because physical parameters can change within a short period of time.
Moreover, because these physical parameters are often numerical values, it is impossible 
to collect a sufficient amount of data for all possible parameters.
Accordingly, we need an unsupervised method that can handle domain shifts while requiring
sound data with only a few sets of physical parameters.

In this paper, we develop an unsupervised anomalous sound detection method
that can handle domain shifts due to changes in physical parameters.
Our idea is to disentangle the factors of domain shifts
and perform anomaly detection by using a space that is invariant with respect to these shifts. 
Specifically, we propose to constrain a neural generative model so that 
some of the latent variables represent factors of domain shifts and 
other latent variables represent components that are invariant with respect to domain shifts.
As a result, the anomaly scores calculated using the latter latent variables are not affected by 
domain shifts but only 
by other variation factors such as a machine's anomalous state, 
which can lead to fewer false positives.\vspace*{-2.8mm}

\section{Problem Statement}\vspace*{-2.8mm}
\label{sec:statement}
Anomalous sound detection is the task of identifying whether a machine
is normal or anomalous according to an anomaly score calculated by a trained model from the machine's sound.
Each piece of input sound data is determined as anomalous if its anomaly score
exceeds a threshold value.
We consider unsupervised anomalous sound detection, in which only normal sound is 
available for training.
We also assume that a machine's physical parameters are only available during training.
This assumption is realistic because sensors to measure the physical parameters may not be available
in real-world operation, depending on environmental conditions.
This problem setting is similar to that of DCASE 2021 Challenge Task 2 \cite{Kawaguchi2021}, 
in which machines have up to three different numerical and physical parameters that cause domain shifts.\vspace*{-2.8mm}

\section{Relation to prior work}\vspace*{-2.8mm}
\subsection{Semi-supervised disentanglement learning}\vspace*{-2.8mm}
Learning disentangled representations has been at the core of representation learning research \cite{Bengio2014}.
Unsupervised disentanglement learning methods, in which 
a VAE with regularizers is commonly used to encourage disentanglement, have mainly been investigated \cite{Higgins2017}.
However, it has been pointed out that unsupervised disentanglement is impossible without 
inductive biases \cite{Locatello2019a}.
Semi-supervised disentanglement learning methods, on the other hand, explicitly use
a few labeled pieces of data to disentangle factors of variation \cite{Locattello2020, Nie2020}.
Locatello et al. \cite{Locattello2020} trained a VAE with an added loss term to incorporate
label information during training.
Esling et al. \cite{Esling2020} used an NF to disentangle categorical tag information from a latent space.
Our proposed method is a semi-supervised disentanglement learning method that uses an
NF to disentangle numerical and physical parameters without an additional loss term. \vspace*{-2.8mm}

\subsection{Anomalous sound detection under domain shifts}\vspace*{-2.8mm}
For DCASE2021 Challenge Task 2, we published the MIMII DUE dataset \cite{Tanabe2021}, the first dataset
for anomalous sound detection under domain shifts. 
In this dataset, we changed physical parameters between the source and target domain
to induce domain shifts. The source domain data and a few samples from the target domain data were available during training.
The top-ranked approaches in the challenge first used autoencoder-based methods or classifiers to extract embeddings
from the data and then used likelihood-based methods like a Gaussian mixture model (GMM) to 
calculate anomaly scores from the embeddings.
Kuroyanagi et al. \cite{Kuroyanagi2021} proposed to calculate anomaly scores by training a GMM for each domain 
on the autoencoder's reconstruction errors.
Sakamoto et al. \cite{Sakamoto2021} attained the highest scores for the target data by 
estimating the mean of the target data under an assumption that the 
mean changes between the source and target domains.
Wilkinghoff \cite{Wilkinghoff2021} trained a classifier that discriminates each set of physical parameters to obtain embeddings.
Our proposed method does not require the target domain data for training, 
as it explicitly uses numerical and physical parameters to obtain a domain-shift-invariant 
latent space that does not change between the source and target domains.\vspace*{-2.8mm}

\section{conventional approach}\vspace*{-2.8mm}
\label{sec:conventional}
\subsection{Semi-supervised disentanglement learning using VAE}\vspace*{-2.8mm}
Locatello et al. \cite{Locattello2020}
proposed a semi-supervised disentanglement learning method to 
disentangle the factors of variations represented by a few labeled data instances, denoted as $\mathbf{y}$. 
They modified the loss function of VAE to incorporate supervision:
\begin{equation}
  \begin{split}
  L = -&\E_{q_\phi (\mathbf{z}|\mathbf{x})} [\log p_\theta (\mathbf{x}|\mathbf{z})] + \beta (D_{KL} (q_\phi (\mathbf{z}|\mathbf{x}))) \\
  &+ \gamma \E_{\mathbf{x}, \mathbf{y}} [R(q_\phi (\mathbf{y}|\mathbf{x}))],
  \end{split}
  \label{Locatello}
\end{equation}
where $\mathbf{x}$ denotes the input data,
$\mathbf{z}$ is the latent variables,
$\beta$ is a hyperparameter introduced in \cite{Higgins2017}, 
$\gamma$ is another hyperparameter,
and $R(\cdot)$ is a function to induce supervised disentanglement.\vspace*{-2.8mm}

\subsection{Unsupervised anomaly detection using NF}\vspace*{-2.8mm}
Among neural generative models for unsupervised anomalous sound detection, 
an NF evaluates the exact likelihood of the input data and 
has shown better detection performance than
other models, including a VAE \cite{Dohi2021}.

The NF models a series of invertible transformations 
$f=f_1 \circ f_2 \circ \cdots \circ f_K$ 
between an input data distribution $p(\mathbf{x})$ and 
a known distribution $p(\mathbf{z})$. 
The log likelihood of the input data can be calculated as
\begin{equation}
  \log p(\mathbf{x}) = \log p(\mathbf{z}_{0}) + \sum_{i=1}^K \log |\det (\frac{d\mathbf{z}_{i}}{d\mathbf{z}_{i-1}})|, 
  \label{likelihood}
\end{equation}
\begin{equation}
  \mathbf{z_i} = f_i (\mathbf{z}_{i-1}),
\end{equation}
where $\mathbf{z_0}$ is a latent vector that follows a known distribution such as the standard isotropic gaussian $N(0, 1)$, 
and $z_i$ $(i=1,2,\cdots,K)$ is an intermediate latent vector. 
The anomaly score can be calculated as the negative 
log likelihood (NLL) of the input data \cite{Schmidt2019, Ryzhikov2019, Dias2020},
\begin{equation}
  a(\mathbf{x}) = -\log p(\mathbf{x}).
  \label{conv_score}
\end{equation}

The NF has mainly been used as an unsupervised method, which fails 
to handle distribution changes due to domain shifts. 
Specifically, when a machine's physical parameters change and the distribution of its 
normal sound data changes, unsupervised methods can output high anomaly scores, leading to false positives.
\vspace*{-2.8mm}

\section{Proposed Approach}\vspace*{-2.8mm}
\label{sec:proposed}
\subsection{Learning of domain-shift-invariant latent space for anomaly detection using NF}\vspace*{-2.8mm}
To handle distribution changes due to domain shifts, we propose to disentangle the factors
of domain shifts and construct a domain-shift-invariant latent space 
for anomaly score calculation.
For this purpose, we use an NF because of its high expressive power.

First, we train an NF model to obtain domain-shift-invariant representations.
As long as the latent variables follow an isotropic Gaussian, if we let some latent variables 
$\mathbf{z}_{d}$ represent a factor of domain shifts, then
the latent space constructed by other latent variables $\mathbf{z}_{c}$ 
should be invariant with respect to that factor.
To make $\mathbf{z}_{d}$ represent a numerical and physical parameter $v$
that causes domain shifts,
we constrain $\mathbf{z}_{d}$ to follow a Gaussian distribution $N(kv, 1)$, where
$k$ is a hyperparameter to induce stable training of the model. 
If a set of sound data with different values of the parameter $v$ is available, then the model can learn
to map input data into a latent space that shifts linearly with $v$.
This forces some latent variables $\mathbf{z}_{d}$ to represent the physical parameter $v$
while making other latent variables $\mathbf{z}_{c}$ invariant to that parameter.

Using the obtained domain shift-invariant latent space, we then calculate anomaly scores
that are unaffected by domain shifts.
The likelihood of the latent variables, $p(\mathbf{z}_{0})$, can be factorized as
\begin{equation}
  \log p(\mathbf{z}_{0}) = \log p(\mathbf{z}_{c}) + \log p(\mathbf{z}_{d}). \label{factor}
\end{equation}
From (\ref{likelihood}) and (\ref{factor}), the likelihood of input data $\mathbf{x}$ can be written as:
\begin{equation}
  \log p(\mathbf{x})
   = \log p(\mathbf{z}_{c}) + \log p(\mathbf{z}_{d}) + \sum_{i=1}^K \log |\det (\frac{d\mathbf{z}_{i}}{d\mathbf{z}_{i-1}})|, \label{data}
\end{equation}
Because the latent variables $\mathbf{z}_{d}$ are constrained to represent 
a factor $v$ of domain shifts, they cannot be used for calculating domain shift-invariant 
anomaly scores.
Also, the third term of (\ref{data}) can remain constant if 
it is calculated using the same learned parameters.
Accordingly, the anomaly score in (\ref{conv_score}) can be calculated
using only the first term of (\ref{data}):
\begin{equation}
  a(\mathbf{z}_{c}) = -\log p(\mathbf{z}_{c}).
  \label{score}
\end{equation}\vspace*{-2.8mm}

\subsection{Multi-scale architecture in NF for learning disentangled reprens\-\-\-entations}\vspace*{-2.8mm}
\label{ssec:multi-scale}
A multi-scale architecture in an NF was first introduced in \cite{Dinh2017} and has since been commonly 
used in other NF models \cite{Kingma2018}.


To apply our method in a multi-scale architecture, we modify only the last block of the 
architecture.
Let $c, h, w$ denote the channel size, height, and width of a feature vector, respectively, and 
assume that the architecture consists of $N$ blocks.
In the $i$th block $(i=1,2,\cdots,N-1)$, an input $\mathbf{x}_{i}$ with dimensions $(c, h, w)$ is squeezed to give 
a feature with dimensions $(4c, h/2, w/2)$.
After some flow transformations, half the channels 
are used as the input $\mathbf{x}_{i+1}$ to the next block, while the other channels, $\mathbf{z}_{ic}$, are factored out; here,
$\mathbf{z}_{ic}$ is constrained to follow $N(0, 1)$.
In the last block, we constrain half the channels, $\mathbf{z}_{d}$, 
to follow $N(0, 1)$ and the others, $\mathbf{z}_{Nc}$, to follow $N(kv, 1)$.
Because only the latent variables in the last block are constrained to represent a factor of domain shifts,
the model has to propagate domain-shift-dependent components to the last block.
As a result, the latent variables factored out at each block can represent domain-shift-invariant components 
at each different scale.
The shift-invariant latent space $\mathbf{z}_{c}$ is obtained by concatenating the $\mathbf{z}_{ic}$ 
 from all the blocks:
\begin{equation}
  \mathbf{z}_{c} = (\mathbf{z}_{1c}, ..., \mathbf{z}_{Nc}).
\end{equation}
The anomaly score is then calculated using (\ref{score}).\vspace*{-2.8mm}

\section{Experiments}\vspace*{-2.8mm}
\label{sec:typestyle}

\subsection{Dataset}
\begin{figure}[t]
  \begin{center}
  \begin{minipage}{0.49\hsize}
  \begin{center}
  \includegraphics[width=1.0\hsize,height=1.8cm,clip]{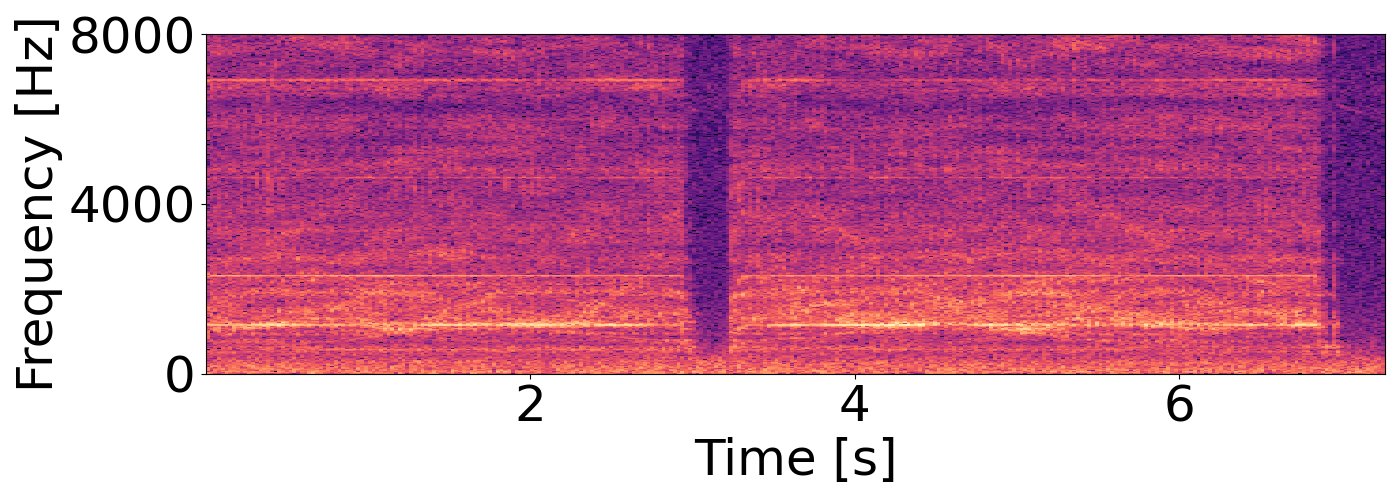}\\
  (a) 100 mm/s\\
  \end{center}
  \end{minipage}
  \begin{minipage}{0.49\hsize}
  \begin{center}
  \includegraphics[width=1.0\hsize,height=1.8cm,clip]{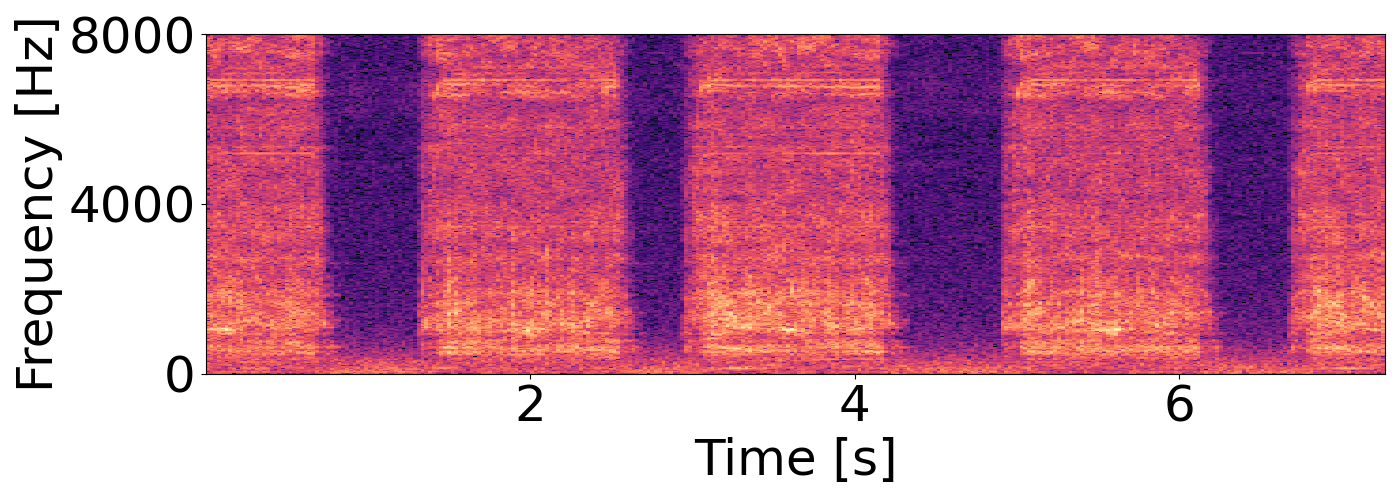}\\
  (b) 300 mm/s\\
  \end{center}
  \end{minipage}
  \begin{minipage}{0.49\hsize}
  \begin{center}
  \includegraphics[width=1.0\hsize,height=1.8cm,clip]{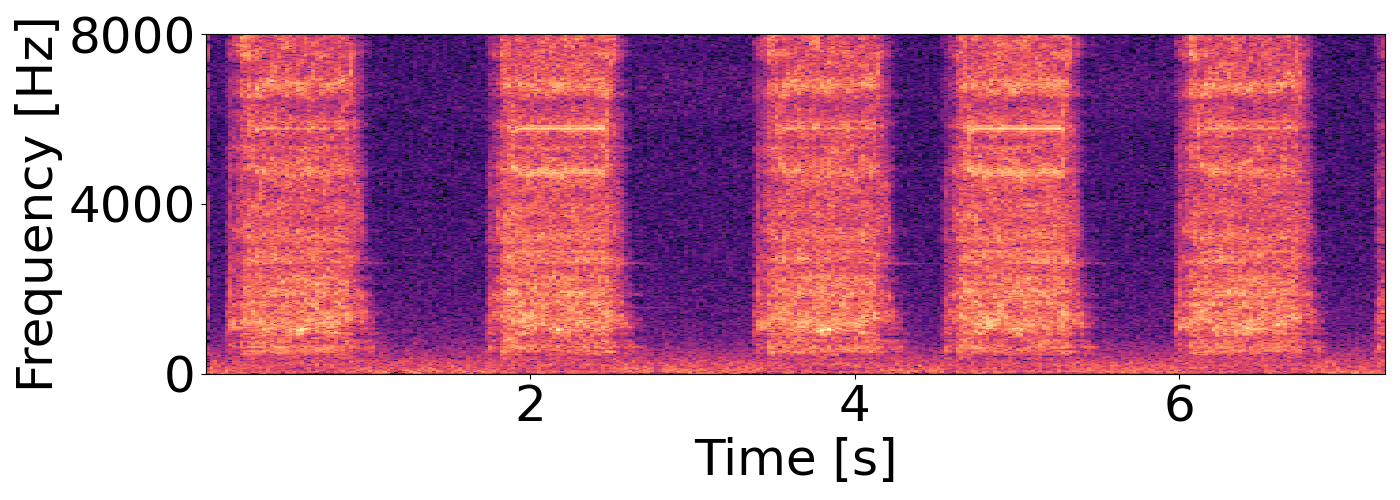}\\
  (c) 500 mm/s\\
  \end{center}
  \end{minipage}
  \begin{minipage}{0.49\hsize}
  \begin{center}
  \includegraphics[width=1.0\hsize,height=1.8cm,clip]{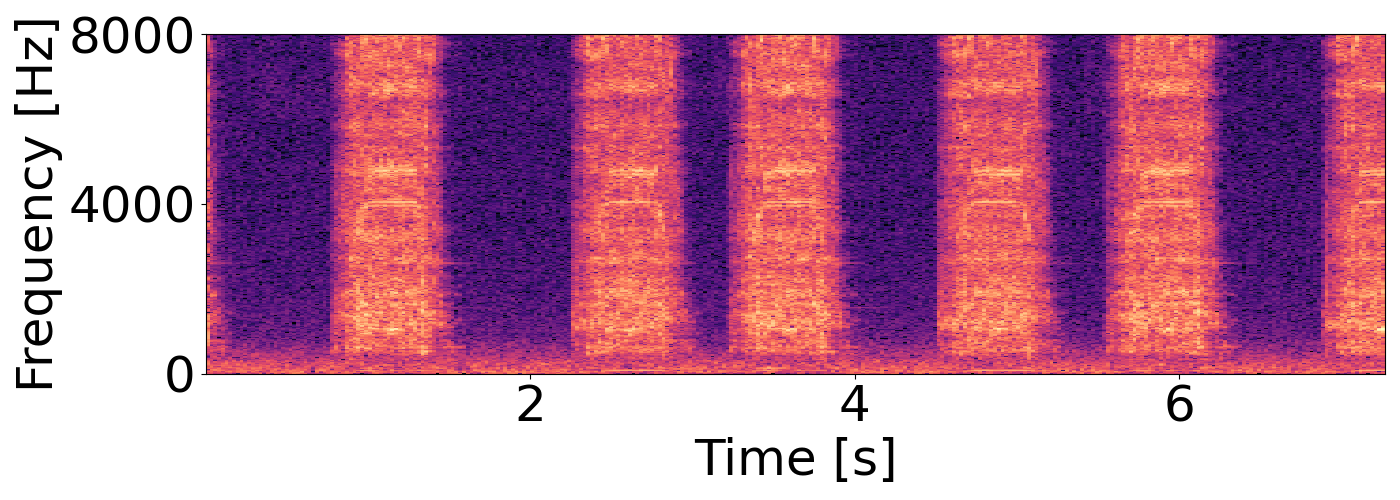}\\
  (d) 700 mm/s\\
  \end{center}
  \end{minipage}
  \caption{Examples of log-mel spectrograms with different operation velocities (mm/s) and 
  an operation distance of 500 mm.}
  \label{fig:spec}
  \end{center}
  \vspace*{-3.5mm}
  \end{figure}

\vspace*{-2.8mm}
We prepared two slide rails with the same machine ID, which operated with a 
designated velocity of 50--750 mm/s and a distance of 100--500 mm.
Figure \ref{fig:spec} shows examples of log-mel spectrograms of the sound with different 
operation velocities.

We first recorded normal sound data with 15 different velocities (50, 100, ..., and 750 mm/s)
and three different distances (100, 250, and 500 mm), giving 45 different 
physical parameter sets in total. Each parameter set had 10 sound clips, with each clip consisting of 
a 10-second single-channel 16-kHz recording.
We used data with different velocities for the training and test data:
seven of the 15 velocities 
(100, 200, 300, 400, 500, 600, and 700 mm/s) were used for the 
training data, while other velocities were used for evaluating the model's ability to disentangle 
velocities that were not included in the training data.

To evaluate the anomaly detection performance, we also r\-e\-c\-o\-r\-d\-e\-d pairs of normal and anomalous sound data.
Both the normal and anomalous sound data had
15 different velocities
and a fixed distance of 500 mm, with 10 sound clips for each velocity.
The normal sound data was recorded using the same slide rail as the one used for the training data.
The anomalous sound data was recorded using the other slide rail.

  \vspace*{-2.8mm}
\subsection{Experimental conditions}\vspace*{-2.8mm}
We used Glow \cite{Kingma2018} for the NF model because it is often chosen 
in out-of-distribution detection and anomaly detection tasks \cite{Dohi2021, Haunschmid2020}.
We also used a VAE with the loss function given in (\ref{Locatello}) to compare the 
disentanglement performance with that of our proposed method.

To obtain input features, we applied the same procedure for Glow and the VAE.
Each frame of the log-mel spectrograms was first computed
with a length of 1024, a hop size of 512, and 128 mel bins. 
At least 313 frames were generated for each sound clip, and 64 consecutive frames
with 48 overlappings were concatenated to generate the input features.

We prepared two Glow models with and without the multi-scale architecture.
The Glow model with the multi-scale architecture (multi-scale Glow) had three blocks with 
five flow steps in each block, and each flow step had 
three CNN layers with 32 hidden layers.
For the Glow model without the multi-scale architecture (single-scale Glow), the only 
difference from the multi-scale Glow was it had just one block.
When using the proposed method, we constrained half the channels of the last block
to follow $N(kv, 1)$, as described in Sec. \ref{ssec:multi-scale}.
In contrast, the single-scale Glow and the multi-scale Glow without the proposed method were trained by 
constraining all the latent variables to follow $N(0, 1)$.

The VAE model had 10 linear layers with 128 dimensions, except for the fifth 
layer, which had eight dimensions. 
The first dimension of the latent variables was constrained to represent 
the velocity $v$, and $R(\cdot)$ in (\ref{Locatello}) was calculated by taking the mean squared error between 
the value of the first latent variable and the actual velocity.
For the hyperparameters we set $\beta=1$ and $\gamma=0.01$.

All models were trained for 1000 epochs by using the 
Adam optimizer \cite{Kingma2014} with a learning rate of 
0.0001 and a batch size of 128.

\begin{figure}[t]
  \centering
    \includegraphics[width=1.0\hsize,height=0.20\vsize,clip]{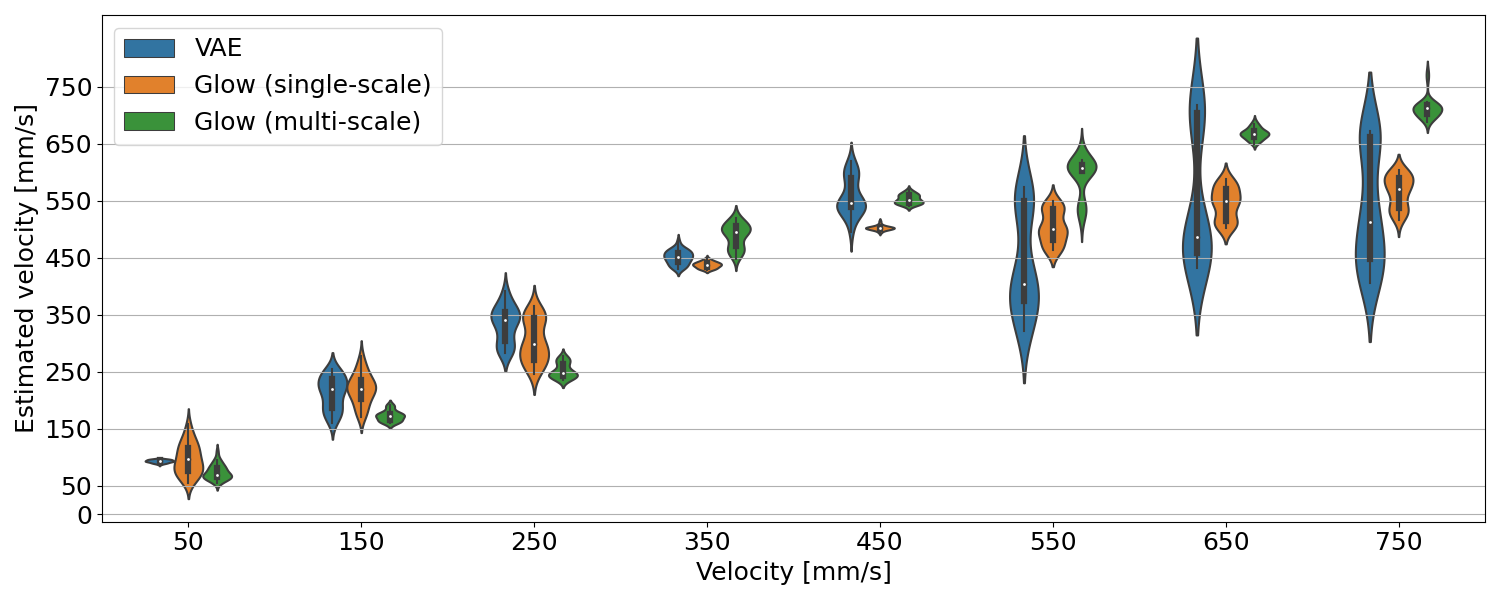}
    \caption{Means of the latent variables for unseen velocities.}
  \label{fig:estimation}
  \vspace*{-5.0mm}
\end{figure}

\begin{figure*}[t]
  \centering
    \includegraphics[width=0.95\hsize,height=0.25\vsize,clip]{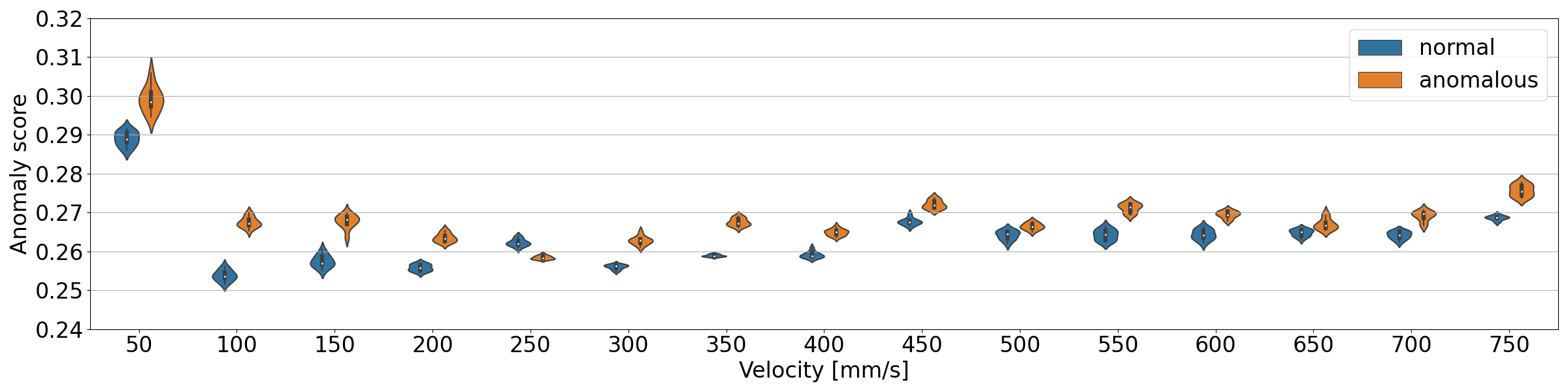}
    \caption{Anomaly scores of sound data with the single-scale Glow and the conventional method.}
  \label{fig:vel0}
  \vspace*{-3.0mm}
\end{figure*}

\begin{figure*}[t]
  \centering
    \includegraphics[width=0.95\hsize,height=0.25\vsize,clip]{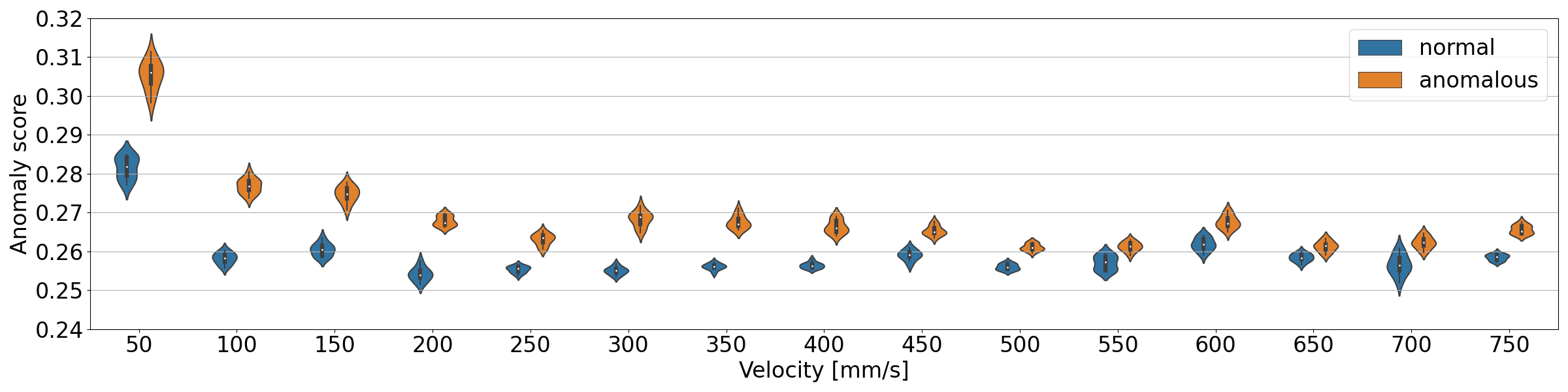}
    \caption{Anomaly scores of sound data with the single-scale Glow and the proposed method.}
  \label{fig:vel1}
  \vspace*{-5.0mm}
\end{figure*}

\vspace*{-2.8mm}
\subsection{Results}\vspace*{-2.8mm}
We first estimated the unseen velocities from the trained models to investigate whether these
velocities could be disentangled.
For both the single-scale and multi-scale Glow, 
the estimated velocity was calculated by taking the mean of $\mathbf{z}_{d}$ in the last block.
For the VAE, the first dimension of the latent variables was used.
The velocities estimated from input features in the same sound clip were averaged to give 
one estimation for each clip.
Figure \ref{fig:estimation} shows the estimation for each unseen velocity.
The multi-scale Glow showed the best performance, especially for the
lower velocities of 50--250 mm/s and the higher velocities of 650 and 750 mm/s.
Though the estimates tended to be larger for the medium velocities of 350--550 mm/s, 
they still showed positive correlations with the actual velocities.
On the other hand, the VAE failed to estimate the higher velocities.
This result shows that the Glow models were better at disentangling the factors of variations
with their higher expressive power.
The multi-scale Glow gave better estimation results than the single-scale Glow, especially for 
higher velocities.
This may be because the multi-scale Glow has more learnable parameters and the 
multi-scale architecture enables extraction of representations at different scales.

\begin{table}[]
  \begin{center}
  \caption{AUCs (in \%) of sound data with the seen velocities (velocities in the training 
  data), unseen velocities (velocities only in the test data), and 
  all velocities.}
  \begin{tabular}{lccc}
  \bhline{1.5pt}
    Method& 
    \begin{tabular}[c]{@{}c@{}}
      Seen\\
      velocities 
    \end{tabular} &
    \begin{tabular}[c]{@{}c@{}}
      Unseen\\
      velocities
    \end{tabular}  & 
    \begin{tabular}[c]{@{}c@{}}
      All\\
      velocities
    \end{tabular}\\ \hline
    \begin{tabular}[c]{@{}l@{}}
      VAE using  \\
      reconstruction error \\
    \end{tabular} & 52.9 & 57.5 & 55.2\\\hline
    \begin{tabular}[c]{@{}l@{}}
      VAE using KLD \\
    \end{tabular} & 47.0 & 52.4 & 49.9\\\hline
    \begin{tabular}[c]{@{}l@{}}
      VAE with loss in (\ref{Locatello}) \\
    \end{tabular} & 64.8 & 60.3 & 62.3\\\hline
    \begin{tabular}[c]{@{}l@{}}
      Single-scale Glow \\
    \end{tabular} & 89.1 & 72.7 & 78.6\\\hline
    \begin{tabular}[c]{@{}l@{}}
      Single-scale Glow using  \\
      proposed method \\
    \end{tabular} & 97.7 & 87.3 & 91.8\\\hline
    \begin{tabular}[c]{@{}l@{}}
      Multi-scale Glow\\
    \end{tabular} & \bf{100.0} & 86.0 & 91.9\\\hline
    \begin{tabular}[c]{@{}l@{}}
      Multi-scale Glow using \\
      proposed method \\
    \end{tabular} & \bf{100.0} & \bf{89.9} & \bf{94.5}\\\bhline{1.5pt}  
  \end{tabular}
  \label{auc}
  \end{center}
  \vspace*{-7.0mm}
\end{table}

We then calculated the anomaly scores 
for each sound clip. 
To evaluate the detection performance of each model, 
we calculated the 
area under the receiver operating characteristic curve (AUC) for the seen velocities, 
unseen velocities, 
and all velocities.
For the VAE, we calculated three different scores: 
the reconstruction error (the first term in (\ref{Locatello})) from a model trained using a conventional loss (first and second terms in (\ref{Locatello})),
the Kullback-Leibler divergence (KLD, second term in (\ref{Locatello})) from the same model, 
and the KLD from a model trained using the loss in (\ref{Locatello}).
For Glow, we used single-scale and multi-scale versions with and without 
the proposed method.
Table \ref{auc} lists the results, which indicate that the proposed method improved the AUC by 
13.2\% for the single-scale Glow and 2.6\% for the multi-scale Glow.
In addition, the Glow models outperformed all the VAE models, 
even though VAE with the loss in (\ref{Locatello}) outperformed 
the other VAEs with the conventional loss. 
The proposed method showed greater improvement in the single-scale Glow
than in the multi-scale Glow. This may be because the number of learnable 
parameters in the single-scale Glow was not enough to learn the 
distribution of the normal data, which made the effect of using the
domain-shift-invariant latent space more evident.

Figures \ref{fig:vel0} and \ref{fig:vel1} show the anomaly scores 
for data using the single-scale Glow with the conventional method and the proposed method, 
respectively.
In Fig. \ref{fig:vel0}, the anomaly scores of the normal sound data with unseen velocities, especially for
50, 250, 450, and 750 mm/s, were higher than those of the normal and even the anomalous sound data 
with seen velocities.
Because the normal data with unseen velocities cannot be used for training, 
the anomaly detector may detect normal samples with these velocities as anomalous samples.
On the other hand, in Fig. \ref{fig:vel1}, the normal sound data with unseen velocities
showed about the same scores as with seen velocities, except for 50 mm/s.
This result indicates that the proposed methods can lower the anomaly scores of normal sound
data with unseen velocities by disentangling the operation velocity.
For 50 mm/s,
because the distribution of the normal data can significantly change around this lower
velocity, the anomaly score for the normal data was higher with this velocity than 
with the other velocities.
The anomaly scores of the anomalous sound data became higher
for several velocities with the proposed method. As the model is 
trained to disentangle velocities by using the normal data, it may have disentangled components that 
that did not represent the velocity of the anomalous sound data, thus raising the anomaly scores.
\vspace*{-2.8mm}
\section{Conclusion}\vspace*{-3.0mm}
We proposed an anomalous sound detection method that can handle domain shifts.
The proposed method uses an NF model to 
disentangle the numerical and physical parameters of a machine, which gives
a domain-shift-invariant latent space.
Experimental results demonstrated that the proposed method disentangles the
factors of domain shifts better than a VAE does, thus enabling improvement in 
the anomaly detection performance
for data with unseen physical parameters.

\clearpage
\ninept
\bibliographystyle{IEEEbib}
\bibliography{refs}

\end{document}